\documentclass[conference]{IEEEtran}
\usepackage{balance}

\ifCLASSINFOpdf
\else
\fi
\usepackage{todonotes}
\usepackage{booktabs}
\usepackage{balance}
\usepackage{amssymb}
\usepackage{pifont}
\usepackage[section]{placeins}

\usepackage{algorithm2e}
\RestyleAlgo{ruled}
\usepackage{listings}
\usepackage{hyperref}
\usepackage{amsmath}
\usepackage{multicol}

\lstdefinelanguage{JavaScript}{
  morekeywords={typeof, new, true, false, catch, function, return, null, catch, switch, var, if, in, while, do, else, case, break},
  morecomment=[s]{/*}{*/},
  morecomment=[l]//,
  morestring=[b]",
  morestring=[b]'
}

\begin{document}
%
\title{Harvest - An Open Source Toolkit for Extracting Posts and Post Metadata from Web Forums}
 
\author{\IEEEauthorblockN{Albert Weichselbraun\IEEEauthorrefmark{1},
Adrian M. P. Brasoveanu\IEEEauthorrefmark{2},
Roger Waldvogel\IEEEauthorrefmark{1},
Fabian Odoni\IEEEauthorrefmark{1}
}
\IEEEauthorblockA{\IEEEauthorrefmark{1}University of Applied Sciences of the Grisons,\\
Pulvermühlestrasse 57, 7000 Chur, Switzerland\\ 
Email: \{firstname.lastname\}@fhgr.ch}
\IEEEauthorblockA{\IEEEauthorrefmark{2}MODUL Technology GmbH\\
Am Kahlenberg 1, 1090 Vienna, Austria\\
Email: adrian.brasoveanu@modul.ac.at
}
}


\IEEEoverridecommandlockouts
\IEEEpubid{\makebox[\columnwidth]{\parbox{\dimexpr\linewidth-2\fboxsep-2\fboxrule}{
  \vspace{1cm}
  © 2020 IEEE. Personal use of this material is permitted. Permission from IEEE must be obtained for all other uses, in any current or future media, including reprinting/republishing this material for advertising or promotional purposes, creating new collective works, for resale or redistribution to servers or lists, or reuse of any copyrighted component of this work in other works.\hfill}}
  \hspace{\columnsep}\makebox[\columnwidth]{ }}

\maketitle
\IEEEpubidadjcol

\begin{abstract}

Automatic extraction of forum posts and metadata is a crucial but challenging task since forums do not expose their content in a standardized structure. Content extraction methods, therefore, often need customizations such as adaptations to page templates and improvements of their extraction code before they can be deployed to new forums. Most of the current solutions are also built for the more general case of content extraction from web pages and lack key features important for understanding forum content such as the identification of author metadata and information on the thread structure. 

This paper, therefore, presents a method that determines the XPath of forum posts, eliminating incorrect mergers and splits of the extracted posts that were common in systems from the previous generation. Based on the individual posts further metadata such as authors, forum URL and structure are extracted. 
We also introduce Harvest, a new open source toolkit that implements the presented methods and create a gold standard extracted from 52 different Web forums for evaluating our approach. A comprehensive evaluation reveals that Harvest clearly outperforms competing systems.
\end{abstract}


\begin{IEEEkeywords}
Information Extraction, Forum Extraction, Natural Language Processing
\end{IEEEkeywords}

%
\IEEEpeerreviewmaketitle

\section{Introduction}\label{sec:introduction}
Forums often contain domain-specific, high quality content that is not available in other sources. In the medical domain, for example, patient forums (e.g., www.healthboards.com and patient.info/forums) contain information on symptoms, drug side effects and patient discussions that offer high value for patient-focused healthcare and drug development. This is particularly true for scarce content such as patient and caretaker discussions on rare diseases that are mostly located in specialized forums.



Automating the extraction of forum posts and metadata is crucial for providing quality social media content extraction services. 
While early solutions involved the use of microformats, a set of JSON or RDF-based data formats for expressing metadata, none of these formats has been widely adopted. In terms of metadata quantity and quality, another serious issue concerns the identification of relevant regions within a forum page (e.g., text, comments, pictures, etc). Web forums might include thread lists, topic-specific entry pages, and multiple pages discussing these topics. Even pages that are focused on discussions can contain threaded lists of text which might be interrupted by images and advertising, therefore, further complicating the content extraction process.
The very limited support for semantic web standards such as microformats, the sheer number of different forum engines used on relevant sites, and the lack of suitable tools have motivated the development of harvest, a toolkit that supports the automatic extraction of content from web forums.

The presented method is primarily focused on the task of forum post and metadata extraction, although it can easily be extended in order to extract additional content such as embedded multimedia files, if needed. The Harvest forum extraction engine described in this paper, the gold standard created for the evaluation as well as the Orbis \cite{DBLP:conf/i-semantics/OdoniKBW18} evaluation plugin used to perform the evaluations from Section~\ref{sec:evaluation} are publicly available\footnote{github.com/fhgr/harvest}.

The main contributions of this paper are
\begin{itemize}
    \item the introduction of a method that extracts forum posts and metadata on the post's date, author, sequence and URL from web forums;
    \item the development of the WEB-FORUM-52 gold standard that contains pages from 52 different web forums used for evaluating the proposed approach; and 
    \item extensive experiments that draw upon multiple evaluation settings to evaluate our method against a baseline and three state of the art content extraction frameworks.
\end{itemize}

The article is organized as follows: Section~\ref{sec:related-work} provides an overview of related work; Section~\ref{sec:method} presents the algorithms we developed for extracting post, date and user information from forum pages; Section~\ref{sec:evaluation} describes the created WEB-FORUMS-52 gold standard and discusses the results of our algorithms. The last section concludes the paper with a set of observations and details on future work.

\section{Related Work}\label{sec:related-work}

Applying machine learning (ML) and deep learning (DL) algorithms to the task of extracting forum content is usually modeled as a multi-step process. At its core, we can identify several large tasks: (i) extraction of the page source code (if needed); (ii) correct identification of all page regions that might contain important data such as navigation, forum posts, user and date information; (iii) extraction of additional information relevant to each of the identified blocks (e.g., description, author metadata, replies to, etc.). 



An early method for extracting forum posts was proposed by Bing Liu in \cite{DBLP:journals/expert/LiuGZ04} and \cite{DBLP:journals/tkde/ZhaiL06}.  The basic idea was to use an unsupervised learning algorithm called Mining Data Region (MDR) which identified data-rich regions from a web page. The method can be applied for different page types, from regular blogs to forums. The basic MDR algorithm is still expensive due to several constraints: (i) it does not work for websites with a flexible structure; (ii) training needs to be performed for each website separately; and (iii) there is no reliable method to separate the various types of data regions from a page (e.g., if a web form contains text advertising, text, comments and annotated comments, all of these will simply be marked as data regions when using MDR). There are multiple variants of MDR including more generalized tree extraction methods like PyDepta \cite{DBLP:journals/tkde/ZhaiL06}, trinary trees \cite{DBLP:journals/tkde/SleimanC14} and template matching \cite{DBLP:journals/tkde/KayedC10}.



Another induction method used to extract user data and a minimal set of metadata is presented in \cite{DBLP:conf/cgc/ZhangJLG12} showing good results. However, this method still requires up to twenty five manual extraction rules. AutoRM \cite{DBLP:journals/kbs/ShiLSYH15} adds a candidate step (Candidate Records) and a filtering step in order to make sure only the interesting candidates are selected. CMDR \cite{wai2017cmdr} uses a Deep Learning node classifier in order to remove the need to train separately for each website. Another article \cite{liu2017deep} goes one step further and uses a convolutional network trained to identify data regions based on their visual properties and uses it further as input for its MDR-like algorithm.

Most of the current ML approaches rely on additional features (e.g., the tag ratios within the HTML document) \cite{DBLP:conf/www/WeningerHH10} and may even require further components such as boilerplate detection \cite{kohlschutter2010boilerplate}) to provide meaningful results. Dragnet \cite{DBLP:conf/www/PetersL13} combines several feature sets into a single library which has demonstrated good results on several early forum extraction datasets. Another recent approach has focused on navigating the hierarchy of objects extracted from web pages (e.g., hyperlink blocks extracted from DOM trees) \cite{DBLP:conf/webi/ZhaoLPBW17}. 
 

Summit Bhatia uses inference networks to extract forum threads \cite{DBLP:conf/aaai/BhatiaM10} and includes a dataset of forum pages collected from Ubuntu and TripAdvisor pages. Some of Bhatia's later articles are focused on forum classification \cite{DBLP:conf/webdb/BhatiaM12}, forum summarization \cite{DBLP:conf/emnlp/BhatiaBM14}, subjectivity detection in forums \cite{DBLP:conf/ecir/BiyaniBCM15}
and detection of factual or discursive threads \cite{DBLP:journals/kbs/BiyaniBCM14}, showcasing the different processing options available for the information extracted from forum data.


In the medical domain extracting forum data has also been approached as a domain-specific problem. This is due to the fact that forums are often one of the few available sources that provide user generated content on rare diseases or drug side-effects. Medical forum thread retrieval includes an additional step of filtering irrelevant information \cite{DBLP:conf/bcb/ChoSZS14}. Some of the recent medical tools have also showcased good results (e.g., the Vigi4Med scraper \cite{audeh2017vigi4med}), but as far as we have noticed, their results and methodology has not been replicated in other domains. PREDOSE \cite{DBLP:journals/jbi/CameronSDSDCACWF13} uses semantic, syntactic and contextual features to enable prescription drug abuse data search through medical forums and social media, however its main focus is on entity and relation extraction, not necessarily full content extraction. Baskaran \cite{baskaran2018automated} presents an automated method of extracting medical information from forums that is based on semantic analysis, but unfortunately difficult to adapt to other domains. 

\section{Method}
\label{sec:method}
Web forums are complex websites that contain a considerable amount of content and metadata such as information on the post's author, date and thread structure. 

From a technical perspective extracting these discussions is challenging since they leverage different blog and forum engines which usually do not have a clear, machine-readable structure and differ significantly in terms of navigation, thread structure and page style, rendering most automatic content extraction attempts ineffective. Forum posts might also contain different types of styling, images and emojis. In addition, post metadata (e.g., user data, date, language) can provide context for subsequent tasks such as forum classification when the post is processed by an automated Natural Language Processing pipeline. 

The discussions in this section elaborate on the following sub-tasks relevant to the presented approach: (i) the identification of the forum posts' XPath (Section~\ref{sec:method-post}) which acts as an anchors for extracting the post content and metadata; and (ii) the subsequent tasks of extracting the post's content (Section~\ref{sec:method-post-content}), date (Section~\ref{sec:method-post-date}), URL (Section~\ref{sec:method-post-link}), and author (Section~\ref{sec:method-post-user}).

\subsection{Identification of forum posts}
\label{sec:method-post}

We identify posts by combining a textual representation of the Web page with information on the document's DOM tree. Our content extraction strategy draws upon the following two observations:
\begin{enumerate}
    \item usually most of the textual content present in Web forums is located in the posts.
    \item the XPath to the forum posts yields multiple sibling nodes, i.e. one sibling per post.
\end{enumerate}

We, therefore, first obtain a textual representation of the page's content (\emph{pageContent}) that has been derived from the HTML to text engine \emph{inscriptis}\footnote{gitlab.com/weblyzard/inscriptis}.  

Algorithm~\ref{alg:post} illustrates how the textual content is then split into lines for which the corresponding XPath is obtained ({\tt getContentXPath}). Afterwards the algorithm computes a score ({\tt xpathScore}; see Algorithm~\ref{alg:post-score}) that indicates the textual coverage of the given XPath. 
Line~\ref{line:constraint} of Algorithm~\ref{alg:post} introduces the constraint that the extracted candidate path needs to yield at least {\tt MIN\_POST\_COUNT} siblings, therefore, ensuring that the extracted structure appears (similar to a forum post) multiple times on the analyzed forum page. Selecting the XPath with the highest {\tt xpathScore} yields the XPath for the forum post. 

\begin{algorithm}
\SetAlgoLined
\LinesNumbered
\KwData{pageContent, domTree}
\KwResult{The XPath to the forum posts}
candidatePaths = [] \;
\ForEach{line in pageContent.split('$\backslash$n')}{
    xpath $\leftarrow$ getContentXPath(domTree, line) \;
    xpathScore $\leftarrow$ getScore(pageContent, xpath, domTree) \;
    xpathElementCount $\leftarrow$ countSiblings(xpath, domTree) \;
    \label{line:constraint}
    \If{(xpathElementCount $>$ MIN\_POST\_COUNT)}{
        candidatePaths.append([xpath, xpathScore, xpathElementCount]) \;
    }
}
\KwRet{getHighestScoringPath(candidatePaths)} \;
\caption{\label{alg:post}Computation of the forum post XPath.}
\end{algorithm}

Algorithm~\ref{alg:post-score} outlines the computation of the similarity metric used for assessing an XPath's coverage of the total page content. The algorithm takes three inputs: (a) a text representation of the page's content ({\tt pageContent}), (b) the XPath to evaluate and (c) the forum's DOM tree. It then computes the cosine similarity to determine the overlap between the forum coverage (i.e. the text present in the nodes that match the provided XPath) and the total page content. In addition, Algorithm~\ref{alg:post-score} introduces constraints on the forum XPath by blacklisting HTML tags such as {\tt <OPTION>, <FOOTER>, <FORM>, <HEAD>}, etc. which clearly indicate that the suggested XPath is unlikely to contain forum posts. Blacklisted HTML tags that are part of an ancestor node yield discounted similarity scores since they only reduce the probability of a forum post.

\begin{algorithm}
\SetAlgoLined
\LinesNumbered
\KwData{pageContent, xpath, domTree}
\KwResult{A score that indicates the node's likelihood of being the post container.}
\tcc{ignore nodes with descendants in the tag blacklist.}
    \If{(decendantsContainBlacklistedTags(domTree, xpath))}{
      return 0.0\;
    }
    \tcc{cosine similarity between the page's content and the text enclosed by the selected xpath.}
    nodeText $\leftarrow$ getNodeTreeText(domTree, xpath)\;
    vsmPageContent $\leftarrow$ getVsm(pageContent)\;
    vsmNode $\leftarrow$ getVsm(nodeText)\;
    sim $\leftarrow$ $\frac{vsmContent \cdot vsmNode}{||vsmContent|| \cdot ||vsmNode||}$ \;

    \tcc{discount nodes with blacklisted ancestors.}
    \If{(ancestorsContainBlacklistedTags(domTree, xpath))}{
        sim $\leftarrow$ sim/10 \;
    }
    \KwRet{sim}\;
\caption{\label{alg:post-score}Evaluation of a node's likelihood of being a post container}
\end{algorithm}

\subsection{Post content}
\label{sec:method-post-content}
Harvest obtains the post content by selecting an ordered list of DOM nodes that match the post XPath. Afterwards, the HTML within each of the selected DOM nodes is converted to the corresponding text yielding the list of forum posts.

\subsection{Date extraction}
\label{sec:method-post-date}

Harvest's data extraction component analyses the DOM tree in the vicinity of the post XPath and identifies candidate dates by locating HTML time elements which contain the \emph{datetime} attribute, and by using dateparser\footnote{github.com/scrapinghub/dateparser} to locate additional candidate elements.

We collect all candidate elements that match the criteria outlined above and then remove dates that are older than 1993-04-30\footnote{The date on which the Internet was opened to the public by CERN.} as well as candidates where the number of extracted dates does not correspond to the number of forum posts. We have relaxed the later constraint to allow for a difference of up to two posts less, since our experiments showed some rare cases where the leading posts used a different layout with another element storing the post's creation date. 

Finally, the candidate elements are scored and sorted according to the following criteria:
\begin{enumerate}
    \item \textit{Post sequence}: XPaths yielding chronologically ascending or descending dates are preferred.
    \item \textit{Recentness}: In most cases, the most recent date refers to the post's creation date. We, therefore, favor XPath candidates returning the most recent dates.
\end{enumerate}
Afterwards, the highest scoring XPath is selected.

\subsection{Post link extraction}
\label{sec:method-post-link}

Harvest locates candidates for the post link extraction by searching for HTML link (\emph{href}) and anchor (\emph{name}) attributes within the individual posts. Only candidates that point to the forum page's URL and yield one link per individual post are considered. In addition, a number is searched at the end of the URL. If this number increases continuously, it is almost certainly a direct link to the post.

\subsection{User extraction}
\label{sec:method-post-user}
Harvest's post user extraction identifies candidates for the post's author by (i) searching link elements other than the post link that do not refer to another web page, and (ii) identifying potential user names in the HTML elements \emph{span}, \emph{strong}, \emph{div} and \emph{b}. We only consider candidates that suggest user names with less than 100 characters and less than four words.

Class attributes containing the words \emph{user}, \emph{member}, \emph{person} or \emph{profile} raise the score of the corresponding XPath. The score also increases if the extracted user name differs between individual posts. In addition, links are weighted higher than text.

Sorting the candidate XPaths based on the obtained score yields the most likely XPath for the post's creator.

%
%
\section{Evaluation}\label{sec:evaluation}


Appropriate benchmarking suites and gold standard data are key towards evaluating content extraction methods, identifying their strengths and weaknesses. We, therefore, have created a gold standard dataset that is used in conjunction with the Open Source Orbis benchmarking framework \cite{odoni_introducing_2019} to evaluate Harvest's performance. 

The evaluation section first provides a short description of the created WEB-FORUM-52 gold standard and then describes the evaluation system.
Afterwards, we introduce two sets of experiments:
\begin{enumerate}
    \item An evaluation of the \emph{post content extraction} which compares Harvest to a baseline and other state-of-the art content extraction frameworks; and
    \item an assessment of Harvest's forum metadata extraction capabilities. 
\end{enumerate}
We conclude the section with a discussion of the evaluation results while taking into account some of the systems that were unavailable for testing in order to provide a clear picture of where Harvest currently stands.

Table~\ref{tab:capabilities} summarizes the capabilities of the evaluated systems. General content extraction systems such as boilerpy and jusText only provide a text representation of the relevant page content. Dragnet and Harvest, in contrast, also extract singular posts and Harvest is the only system capable of extracting post metadata as well.
Inscriptis acts as a baseline since it has been designed for converting HTML to text and, therefore, also returns boilerplate elements such as navigation areas and copyright notes.

\begin{table}[htb]
\caption{\label{tab:capabilities}Capabilities of the evaluated systems. Text refers to the extraction of the forum's text, post to the identification of individual posts and date, user, URL to the extraction of the corresponding metadata.}
\centering
\begin{tabular}{llllll}
\toprule
System     & text           & post & date & user & URL \\ \midrule
Inscriptis (baseline)&  $\checkmark$  &      &      &      &     \\
boilerpy   &  $\checkmark$  &      &      &      &     \\
Dragnet    &  $\checkmark$  & $\checkmark$     &      &      &     \\
jusText    &  $\checkmark$  &     &      &      &     \\
Harvest    &  $\checkmark$  & $\checkmark$     & $\checkmark$  & $\checkmark$ &    $\checkmark$ \\ \bottomrule
\end{tabular}
\end{table}

\subsection{The WEB-FORUM-52 gold standard} 
\label{sec:gold-standard}

The WEB-FORUM-52 gold standard comprises (i) 13 web forums from the health domain, (ii) 15 forums obtained from a Wikipedia list of popular forums\footnote{en.wikipedia.org/wiki/List\_of\_Internet\_forums}, (iii) 13 forums mentioned on a list of popular German Web forums\footnote{www.beliebte-foren.de/}, (iv) nine forums obtained from WPressBlog\footnote{www.wpressblog.com/free-forum-posting-sites-list/} and (v) two additional forums.

For most forums two web pages (from different threads) were used and stored together with gold standard annotations that have been manually created by domain experts and describe the post text, post date, post user and direct URL to the post. 



The gold standard is publicly available on GitHub\footnote{github.com/fhgr/harvest}.

\subsection{Evaluation system}

Since one of our long-term goals when creating new experiments is to enhance transparency and reproducibility, we have decided to use an open source framework for computing the evaluation scores. We have selected Orbis \cite{DBLP:conf/i-semantics/OdoniKBW18} which was designed with extensibility in mind.  Although Orbis also enables visual evaluations, we created a forum-extraction evaluation plugin that solely uses the Orbis command line interface, since designing new visual evaluations is beyond the scope of this work. 

Barbaresi and Lejeune \cite{DBLP:conf/aclwac/BarbaresiL20} present an extensive evaluation of the best content extraction tools available in early 2020. In this evaluation Inscriptis\footnote{github.com/weblyzard/inscriptis} proved to be the fastest tool and also yielded the best recall. Dragnet\footnote{github.com/dragnet-org/dragnet} in turn, provided the highest precision, and depending on the metric (e.g., clean-eval, euclidean or cosine distances, etc) either Dragnet or News-Please\footnote{github.com/fhamborg/news-please} yielded the best F1 score. For our experiments, we have selected several of the systems used in Barberesi's evaluation \cite{DBLP:conf/aclwac/BarbaresiL20} including Inscriptis, BoilerPy3\footnote{github.com/jmriebold/BoilerPy3}, jusText\footnote{github.com/miso-belica/jusText} and Dragnet. Several other tools have been initially targeted but were not included since their source code was not available online at the date the experiments were performed (e.g., Sido's forum extraction tool \cite{DBLP:journals/cys/SidoKP19}) or due to various errors (e.g., the News-Please tool). We will continue to pursue the developers of these tools to include their systems in future versions of the created evaluation plugin.

\subsection{Post content extraction}
\label{sec:experiments}
This experiment evaluates how well the extracted post content corresponds to the gold standard data. Harvest is compared to three other content extraction methods and to a baseline (Inscriptis) that extracts the whole text from the Web page (i.e. forum posts and boilerplate content). 

Most of the forum extraction evaluations tend to be recall-oriented, therefore the classic metrics used within them are recall at various cutoffs (R@N) and Mean Reciprocal Rank (MRR) which evaluates a list of possible responses to a set of sample queries and generally makes sense to use when only a single relevant document is known (e.g., one relevant page from a forum). Recall-oriented evaluations are rather well-suited for settings in which pages are extracted from a single forum, and, therefore, have not been considered in the present evaluation. 

In precision-oriented settings (e.g., production environments), it is also customary to compute precision at various cutoffs (e.g., P@N) and Mean Average Precision (mAP). Depending on the algorithms that are evaluated (e.g., MDR algorithms, clustering algorithms), some forum evaluations have also provided additional metrics like Adjusted Rand Index (ARI) and Adjusted Mutual Information (AMI), measures that are used for establishing the similarity between clustering algorithms.

Since we were interested in comparing the actual text of the posts returned by each tool, we have considered the following performance metrics that are tailored towards evaluating content extraction tools: (a) Levenshtein distance \cite{levenshtein_binary_1966}, (b) the Jaccard Coefficient, and (c) a token-based computation of precision, recall, and the F1 measure \cite{weninger_cetr_2010}.

The equations below use the following notation: $S_g$ refers to the string containing the gold standard text, $S_e$ to the string that has been extracted by the evaluated systems. For token-based measures these strings are split into tokens $t_i$ yielding two token sets - one for the gold standard ($T_g$) and a second one for the extracted text ($T_e$).

\subsubsection{Levenshtein distance}
The first approach computes the normalized Levenshtein distance ($lev_{norm}$) between the gold standard forum text ($S_g$) and the forum text extracted by the systems ($S_e$)

\begin{eqnarray}
  lev_{norm}(S_g, S_e) &=& \frac{lev(S_g, S_e)}{max(|S_g|, |S_s|)}
\end{eqnarray}
where $|S_g|$ and $|S_e|$ refer to the length of the gold standard and extracted text respectively.
The evaluated systems often halved, split or incorrectly merged posts which seriously impacts the time required for computing the Levenshtein distance. We, therefore, selected the FuzzyWuzzy Python package\footnote{github.com/seatgeek/fuzzywuzzy} which provides a fast computation of the Levenshtein distance \cite{levenshtein_binary_1966} even under the stated conditions.

\subsubsection{Jaccard Coefficient}
A second similarity measure is the Jaccard Coefficient  which is computed based on the extracted token sets as outlined below:

\begin{eqnarray}
  J(T_g, T_e) &=& |T_g \cap T_e|/|T_g \cup T_e|
\end{eqnarray}


\subsubsection{Token-based similarity}
The token-based similarity metric computes precision, recall and consequently the F1 measure based on the common tokens between the gold standard and the extracted text.
\begin{multicols}{2}
\noindent
\begin{eqnarray}
   P &=& \frac{|T_g \cap T_e|}{T_e}
\end{eqnarray}
\begin{eqnarray}
   R &=& \frac{|T_g \cap T_e|}{T_g}
\end{eqnarray}
\end{multicols}

We compute both micro (mP, mR and mF1) and macro (MP, MR and MF1) results for the evaluations. The micro results correspond to the weighted average scores, whereas the macro results represent the arithmetic mean of the per-class (e.g., type) scores. Micro results are well-suited for evaluating results of imbalanced classes, whereas the macro-averages compute the metrics separately for each class and then take the averages. It is important to provide both metrics precisely because class distributions differ wildly between various pages or corpora.

\begin{table}
  \caption{Evaluation of the post extraction task using Levenshtein distance. Micro and Macro precision, recall and F1 scores.}
  \label{tab:summary-levenshtein}
  \begin{tabular}{lllllll}
  \toprule
  Method & mP & mR & mF1 & MP & MR & MF1 \\
  \midrule
  Dragnet 2.0.4 & 0.26 & 0.47 & 0.33 & 0.35 & 0.48 & 0.37 \\
  jusText 2.2.0 & 0.73 & 0.63 & 0.68 & 0.63 & 0.63 & 0.63 \\
  BoilerPy3 1.0.2 & 0.50 & 0.49 & 0.50 & 0.49 & 0.49 & 0.49 \\
  Inscriptis 1.1.0 & 0.37 & 0.37 & 0.37 & 0.37 & 0.37 & 0.37 \\
  Harvest 1.0.0 & 0.93 & 0.87 & 0.90 & 0.86 & 0.86 & 0.86 \\\bottomrule
  \end{tabular}
\end{table}

\begin{table}
  \caption{Evaluation of the post extraction task using Jaccard Coefficient. Micro and Macro precision, recall and F1 scores.}
  \label{tab:summary-jaccard}
  \begin{tabular}{lllllll}
  \toprule
  Method & mP & mR & mF1 & MP & MR & MF1 \\
  \midrule
  Dragnet 2.0.4 & 0.24 & 0.43 & 0.3 & 0.33 & 0.45 & 0.34 \\
  jusText 2.2.0 & 0.52 & 0.45 & 0.48 & 0.45 & 0.45 & 0.45 \\
  BoilerPy3 1.0.2 & 0.30 & 0.29 & 0.30 & 0.29 & 0.29 & 0.29 \\
  Inscriptis 1.1.0 & 0.31 & 0.31 & 0.31 & 0.31 & 0.31 & 0.31 \\
  Harvest 1.0.0 & 0.92 & 0.87 & 0.89 & 0.86 & 0.85 & 0.85 \\\bottomrule
  \end{tabular}
\end{table}

\begin{table}
  \caption{Micro and Macro precision, recall and F1 scores. Precision, recall and F1 computed with the approach from  Weninger et al. }
  \label{tab:summary-weniger}
  \begin{tabular}{lllllll}
  \toprule
  Method & mP & mR & mF1 & MP & MR & MF1 \\
  \midrule
  Dragnet 2.0.4 & 0.94 & 0.49 & 0.65 & 0.87 & 0.64 & 0.70 \\
  jusText 2.2.0 & 0.93 & 0.73 & 0.82 & 0.78 & 0.60 & 0.66 \\
  BoilerPy3 1.0.2 & 0.95 & 0.50 & 0.65 & 0.83 & 0.47 & 0.57 \\
  Inscriptis 1.1.0 & 0.71 & 0.99 & 0.83 & 0.34 & 0.55 & 0.41 \\
  Harvest 1.0.0 & 0.99 & 0.99 & 0.99 & 0.91 & 0.91 & 0.91 \\\bottomrule
  \end{tabular}
\end{table}

\begin{table}[htb]
    \centering
    \caption{Evaluation of Harvest's 1.0.0 metadata extraction performance.}
    \label{tab:post-metadata}
\begin{tabular}{@{}lllllll@{}}\toprule
Metadata field  & mP & mR & mF1 & MP & MR & MF1\\\midrule
  post user     & 0.86 & 0.79 & 0.83 & 0.76 & 0.76 & 0.76 \\
  post date     & 0.41 & 0.33 & 0.36 & 0.37 & 0.38 & 0.38 \\
  post URL      & 0.51 & 0.66 & 0.58 & 0.43 & 0.42 & 0.42 \\ \bottomrule
\end{tabular}
\end{table}

\subsection{Experimental results}

Tables~\ref{tab:summary-levenshtein}, \ref{tab:summary-jaccard} and \ref{tab:summary-weniger} summarize the evaluation results. Regardless of the performance metric used, Harvest provided the best performance for the post extraction tasks. Since some of the tools included in this evaluation were primarily designed for general content extraction (e.g., Dragnet), a wrapper was added to correct mistakes in the partitioning of the posts. Example of such errors include: (i) the merging of multiple forum posts into a single post, as most tools will definitely end up with a variant of this issue which needs to be dealt with (e.g., merging of consecutive posts, merging of final post in a page with the footer of the page); (ii) the reverse error - which is splitting a post into multiple posts (e.g., due to pictures or other media, posts can sometimes end up being broken into multiple pieces); or (iii) merging of post metadata and post content (e.g., instead of extracting user metadata and post content into separate slots, they end up all merged into the content slot). 
The wrappers draw upon lists of conversation starters and enders in several languages (e.g., English, German, Spanish), to correct the partitioning obtained from the original tools. 

Besides the three error classes mentioned above, we have also encountered error classes that were less frequent or specific to particular tools. 

Dragnet, for example, has repeatedly lost various posts from multiple forum websites (e.g., instead of retrieving the content for ten posts from a page, it only retrieved the content for eight or nine posts) or sometimes returned the content only partially (e.g., retrieved only a sentence or several sentences from a post, but not the full content of the post). The missing posts could be related to the various features used for training Dragnet, but it is currently difficult to understand which features have led to this outcome. 

Justext has performed well in situations in which the post segmentation was clear, but has often failed to correctly separate the posts when media objects were included in the pages (e.g., when lots of adverts or different sections of a page intermingled). 

Boilerpy has sometimes retrieved only the titles or the first forum posts from a page and has barely retrieved any content when the page layouts were similar to blog posts with a larger content section and many comments. Also, similar to Dragnet, in some cases it has randomly lost forum posts. 

Inscriptis is typically used for extracting the HTML and DOM content of a page, therefore some of its errors were related to these use cases. 

Harvest also loses some posts due to page layout (e.g., in several cases it fails to recognize the first post on a page), but generally has less errors due to the content. Most of the Harvest errors currently are due to the failure to retrieve some of the post URLs or user names correctly, but these are currently fixed. However, compared to the rest of the competitors, Harvest does manage to extract almost all of the content available on a forum page.

The second set of experiments (see Table~\ref{tab:post-metadata}) evaluates the performance of Harvest's metadata extraction components which also extracts the post's date, user and link. 
From these three tasks the post user extraction is the most reliable one. The results also clearly indicate that the extraction of the post creation data is currently the most challenging task since dates are not always organized in a separate XPath which requires the combination of multiple tools -- i.e., Harvest for computing the XPath and dateparser for extracting the date. Consequently, even smaller mistakes multiply and reduce the overall effectiveness of the extraction task. 

\section{Outlook and Conclusions}\label{sec:conclusion}

Most content extraction frameworks are generic enough to work in multiple settings (e.g., crawlers that can extract content from classic web pages and social media), but rarely provide good results for custom scenarios such as the extraction of Web  forum posts. This observation has also been confirmed by the evaluation results presented in Section~\ref{sec:evaluation} which indicate that Harvest clearly outperformed other systems for the given forum extraction task. 

Consequently, users can either adapt existing tools such as Dragnet to their tasks or draw upon domain-specific content extractors such as the Vigil4Med scraper, if they are available for the targeted application domain. Harvest although limited to forum extraction, in contrast, addresses both needs: (i) generality and (ii) domain-specificity, therefore, simplifying the task of extracting and processing information from web forums.

Since Harvest focuses on extracting the entire forum content (i.e. high recall) while maintaining a high precision, our evaluations draws upon the F1 metric. Other measures such as ARI, AMI, mAP or P@n are also frequently used in literature. We plan to extend the created Orbis forum extraction evaluation plugin with some of these metrics and also aim to include additional evaluation types that cover an even larger array of content extraction issues in future work.

Our efforts ultimately aim at correctly classifying information from the extracted text based on multiple features like sentiment, entities and symptoms, but during early evaluations we discovered that content extraction is rarely perfect. Future work will return to this original goal and focus mostly on classification and related tasks like joint intent detection and slot filling.

\subsection*{Acknowledgement}
The research presented in this paper has been conducted within the \emph{MedMon} project (www.fhgr.ch/medmon) funded by Innosuisse (No. 25587.2 PFES-ES) and the Gentio (https://www.gentio.eu/) project funded by the Austrian Federal Ministry for Climate Action, Environment, Energy, Mobility and Technology (BMK) via the ICT of the Future Program (GA No. 873992).

\bibliographystyle{IEEEtran}
\balance
\bibliography{forum}

\begin{thebibliography}{10}
\providecommand{\url}[1]{#1}
\csname url@samestyle\endcsname
\providecommand{\newblock}{\relax}
\providecommand{\bibinfo}[2]{#2}
\providecommand{\BIBentrySTDinterwordspacing}{\spaceskip=0pt\relax}
\providecommand{\BIBentryALTinterwordstretchfactor}{4}
\providecommand{\BIBentryALTinterwordspacing}{\spaceskip=\fontdimen2\font plus
\BIBentryALTinterwordstretchfactor\fontdimen3\font minus
  \fontdimen4\font\relax}
\providecommand{\BIBforeignlanguage}[2]{{%
\expandafter\ifx\csname l@#1\endcsname\relax
\typeout{** WARNING: IEEEtran.bst: No hyphenation pattern has been}%
\typeout{** loaded for the language `#1'. Using the pattern for}%
\typeout{** the default language instead.}%
\else
\language=\csname l@#1\endcsname
\fi
#2}}
\providecommand{\BIBdecl}{\relax}
\BIBdecl

\bibitem{DBLP:conf/i-semantics/OdoniKBW18}
\BIBentryALTinterwordspacing
F.~Odoni, P.~Kuntschik, A.~M.~P. Brasoveanu, and A.~Weichselbraun, ``On the
  importance of drill-down analysis for assessing gold standards and named
  entity linking performance,'' in \emph{Proceedings of the 14th International
  Conference on Semantic Systems, {SEMANTICS} 2018, Vienna, Austria, September
  10-13, 2018}, ser. Procedia Computer Science, A.~Fensel, V.~de~Boer,
  T.~Pellegrini, E.~Kiesling, B.~Haslhofer, L.~Hollink, and A.~Schindler, Eds.,
  vol. 137.\hskip 1em plus 0.5em minus 0.4em\relax Elsevier, 2018, pp. 33--42.
  [Online]. Available: \url{https://doi.org/10.1016/j.procs.2018.09.004}
\BIBentrySTDinterwordspacing

\bibitem{DBLP:journals/expert/LiuGZ04}
\BIBentryALTinterwordspacing
B.~Liu, R.~L. Grossman, and Y.~Zhai, ``Mining web pages for data records,''
  \emph{{IEEE} Intell. Syst.}, vol.~19, no.~6, pp. 49--55, 2004. [Online].
  Available: \url{https://doi.org/10.1109/MIS.2004.68}
\BIBentrySTDinterwordspacing

\bibitem{DBLP:journals/tkde/ZhaiL06}
\BIBentryALTinterwordspacing
Y.~Zhai and B.~Liu, ``Structured data extraction from the web based on partial
  tree alignment,'' \emph{{IEEE} Trans. Knowl. Data Eng.}, vol.~18, no.~12, pp.
  1614--1628, 2006. [Online]. Available:
  \url{https://doi.org/10.1109/TKDE.2006.197}
\BIBentrySTDinterwordspacing

\bibitem{DBLP:journals/tkde/SleimanC14}
\BIBentryALTinterwordspacing
H.~A. Sleiman and R.~Corchuelo, ``Trinity: On using trinary trees for
  unsupervised web data extraction,'' \emph{{IEEE} Trans. Knowl. Data Eng.},
  vol.~26, no.~6, pp. 1544--1556, 2014. [Online]. Available:
  \url{https://doi.org/10.1109/TKDE.2013.161}
\BIBentrySTDinterwordspacing

\bibitem{DBLP:journals/tkde/KayedC10}
\BIBentryALTinterwordspacing
M.~Kayed and C.~Chang, ``Fivatech: Page-level web data extraction from template
  pages,'' \emph{{IEEE} Trans. Knowl. Data Eng.}, vol.~22, no.~2, pp. 249--263,
  2010. [Online]. Available: \url{https://doi.org/10.1109/TKDE.2009.82}
\BIBentrySTDinterwordspacing

\bibitem{DBLP:conf/cgc/ZhangJLG12}
\BIBentryALTinterwordspacing
J.~Zhang, C.~Jin, Y.~Lin, and X.~Gong, ``Forum data extraction without explicit
  rules,'' in \emph{2012 Second International Conference on Cloud and Green
  Computing, {CGC} 2012, Xiangtan, Hunan, China, November 1-3, 2012}, J.~Liu,
  J.~Chen, and G.~Xu, Eds.\hskip 1em plus 0.5em minus 0.4em\relax {IEEE}
  Computer Society, 2012, pp. 460--465. [Online]. Available:
  \url{https://doi.org/10.1109/CGC.2012.72}
\BIBentrySTDinterwordspacing

\bibitem{DBLP:journals/kbs/ShiLSYH15}
\BIBentryALTinterwordspacing
S.~Shi, C.~Liu, Y.~Shen, C.~Yuan, and Y.~Huang, ``Autorm: An effective approach
  for automatic web data record mining,'' \emph{Knowl. Based Syst.}, vol.~89,
  pp. 314--331, 2015. [Online]. Available:
  \url{https://doi.org/10.1016/j.knosys.2015.07.012}
\BIBentrySTDinterwordspacing

\bibitem{wai2017cmdr}
F.~K. Wai, L.~W. Yong, V.~L. Thing, and V.~Pomponiu, ``Cmdr: Classifying nodes
  for mining data records with different html structures,'' in \emph{TENCON
  2017-2017 IEEE Region 10 Conference}.\hskip 1em plus 0.5em minus 0.4em\relax
  IEEE, 2017, pp. 1862--1862.

\bibitem{liu2017deep}
J.~Liu, L.~Lin, Z.~Cai, J.~Wang, and H.-j. Kim, ``Deep web data extraction
  based on visual information processing,'' \emph{Journal of Ambient
  Intelligence and Humanized Computing}, pp. 1--11, 2017.

\bibitem{DBLP:conf/www/WeningerHH10}
\BIBentryALTinterwordspacing
T.~Weninger, W.~H. Hsu, and J.~Han, ``{CETR:} content extraction via tag
  ratios,'' in \emph{Proceedings of the 19th International Conference on World
  Wide Web, {WWW} 2010, Raleigh, North Carolina, USA, April 26-30, 2010},
  M.~Rappa, P.~Jones, J.~Freire, and S.~Chakrabarti, Eds.\hskip 1em plus 0.5em
  minus 0.4em\relax {ACM}, 2010, pp. 971--980. [Online]. Available:
  \url{https://doi.org/10.1145/1772690.1772789}
\BIBentrySTDinterwordspacing

\bibitem{kohlschutter2010boilerplate}
C.~Kohlsch{\"u}tter, P.~Fankhauser, and W.~Nejdl, ``Boilerplate detection using
  shallow text features,'' in \emph{Proceedings of the third ACM international
  conference on Web search and data mining}, 2010, pp. 441--450.

\bibitem{DBLP:conf/www/PetersL13}
\BIBentryALTinterwordspacing
M.~E. Peters and D.~Lecocq, ``Content extraction using diverse feature sets,''
  in \emph{22nd International World Wide Web Conference, {WWW} '13, Rio de
  Janeiro, Brazil, May 13-17, 2013, Companion Volume}, L.~Carr, A.~H.~F.
  Laender, B.~F. L{\'{o}}scio, I.~King, M.~Fontoura, D.~Vrandecic, L.~Aroyo,
  J.~{Palazzo M. de Oliveira}, F.~Lima, and E.~Wilde, Eds.\hskip 1em plus 0.5em
  minus 0.4em\relax International World Wide Web Conferences Steering Committee
  / {ACM}, 2013, pp. 89--90. [Online]. Available:
  \url{https://doi.org/10.1145/2487788.2487828}
\BIBentrySTDinterwordspacing

\bibitem{DBLP:conf/webi/ZhaoLPBW17}
\BIBentryALTinterwordspacing
K.~Zhao, B.~Li, Z.~Peng, J.~Bu, and C.~Wang, ``Navigation objects extraction
  for better content structure understanding,'' in \emph{Proceedings of the
  International Conference on Web Intelligence, Leipzig, Germany, August 23-26,
  2017}, A.~P. Sheth, A.~Ngonga, Y.~Wang, E.~Chang, D.~Slezak, B.~Franczyk,
  R.~Alt, X.~Tao, and R.~Unland, Eds.\hskip 1em plus 0.5em minus 0.4em\relax
  {ACM}, 2017, pp. 629--636. [Online]. Available:
  \url{https://doi.org/10.1145/3106426.3106437}
\BIBentrySTDinterwordspacing

\bibitem{DBLP:conf/aaai/BhatiaM10}
\BIBentryALTinterwordspacing
S.~Bhatia and P.~Mitra, ``Adopting inference networks for online thread
  retrieval,'' in \emph{Proceedings of the Twenty-Fourth {AAAI} Conference on
  Artificial Intelligence, {AAAI} 2010, Atlanta, Georgia, USA, July 11-15,
  2010}, M.~Fox and D.~Poole, Eds.\hskip 1em plus 0.5em minus 0.4em\relax
  {AAAI} Press, 2010. [Online]. Available:
  \url{http://www.aaai.org/ocs/index.php/AAAI/AAAI10/paper/view/1886}
\BIBentrySTDinterwordspacing

\bibitem{DBLP:conf/webdb/BhatiaM12}
\BIBentryALTinterwordspacing
S.~Bhatia, P.~Biyani, and P.~Mitra, ``Classifying user messages for managing
  web forum data,'' in \emph{Proceedings of the 15th International Workshop on
  the Web and Databases 2012, WebDB 2012, Scottsdale, AZ, USA, May 20, 2012},
  Z.~G. Ives and Y.~Velegrakis, Eds., 2012, pp. 13--18. [Online]. Available:
  \url{http://db.disi.unitn.eu/pages/WebDB2012/papers/p26.pdf}
\BIBentrySTDinterwordspacing

\bibitem{DBLP:conf/emnlp/BhatiaBM14}
\BIBentryALTinterwordspacing
------, ``Summarizing online forum discussions - can dialog acts of individual
  messages help?'' in \emph{Proceedings of the 2014 Conference on Empirical
  Methods in Natural Language Processing, {EMNLP} 2014, October 25-29, 2014,
  Doha, Qatar, {A} meeting of SIGDAT, a Special Interest Group of the {ACL}},
  A.~Moschitti, B.~Pang, and W.~Daelemans, Eds.\hskip 1em plus 0.5em minus
  0.4em\relax {ACL}, 2014, pp. 2127--2131. [Online]. Available:
  \url{https://doi.org/10.3115/v1/d14-1226}
\BIBentrySTDinterwordspacing

\bibitem{DBLP:conf/ecir/BiyaniBCM15}
\BIBentryALTinterwordspacing
P.~Biyani, S.~Bhatia, C.~Caragea, and P.~Mitra, ``Using subjectivity analysis
  to improve thread retrieval in online forums,'' in \emph{Advances in
  Information Retrieval - 37th European Conference on {IR} Research, {ECIR}
  2015, Vienna, Austria, March 29 - April 2, 2015. Proceedings}, ser. Lecture
  Notes in Computer Science, A.~Hanbury, G.~Kazai, A.~Rauber, and N.~Fuhr,
  Eds., vol. 9022, 2015, pp. 495--500. [Online]. Available:
  \url{https://doi.org/10.1007/978-3-319-16354-3\_54}
\BIBentrySTDinterwordspacing

\bibitem{DBLP:journals/kbs/BiyaniBCM14}
\BIBentryALTinterwordspacing
------, ``Using non-lexical features for identifying factual and opinionative
  threads in online forums,'' \emph{Knowl. Based Syst.}, vol.~69, pp. 170--178,
  2014. [Online]. Available: \url{https://doi.org/10.1016/j.knosys.2014.04.048}
\BIBentrySTDinterwordspacing

\bibitem{DBLP:conf/bcb/ChoSZS14}
\BIBentryALTinterwordspacing
J.~H.~D. Cho, P.~Sondhi, C.~Zhai, and B.~R. Schatz, ``Resolving healthcare
  forum posts via similar thread retrieval,'' in \emph{Proceedings of the 5th
  {ACM} Conference on Bioinformatics, Computational Biology, and Health
  Informatics, {BCB} '14, Newport Beach, California, USA, September 20-23,
  2014}, P.~Baldi and W.~Wang, Eds.\hskip 1em plus 0.5em minus 0.4em\relax
  {ACM}, 2014, pp. 33--42. [Online]. Available:
  \url{https://doi.org/10.1145/2649387.2649399}
\BIBentrySTDinterwordspacing

\bibitem{audeh2017vigi4med}
B.~Audeh, M.~Beigbeder, A.~Zimmermann, P.~Jaillon, and C.~Bousquet, ``Vigi4med
  scraper: A framework for web forum structured data extraction and semantic
  representation,'' \emph{PloS one}, vol.~12, no.~1, 2017.

\bibitem{DBLP:journals/jbi/CameronSDSDCACWF13}
\BIBentryALTinterwordspacing
D.~Cameron, G.~A. Smith, R.~Daniulaityte, A.~P. Sheth, D.~Dave, L.~Chen,
  G.~Anand, R.~Carlson, K.~Z. Watkins, and R.~Falck, ``{PREDOSE:} {A} semantic
  web platform for drug abuse epidemiology using social media,'' \emph{J.
  Biomed. Informatics}, vol.~46, no.~6, pp. 985--997, 2013. [Online].
  Available: \url{https://doi.org/10.1016/j.jbi.2013.07.007}
\BIBentrySTDinterwordspacing

\bibitem{baskaran2018automated}
\BIBentryALTinterwordspacing
U.~Baskaran and K.~Ramanujam, ``Automated scraping of structured data records
  from health discussion forums using semantic analysis,'' \emph{Informatics in
  Medicine Unlocked}, vol.~10, pp. 149--158, 2018. [Online]. Available:
  \url{https://doi.org/10.1016/j.imu.2018.01.003}
\BIBentrySTDinterwordspacing

\bibitem{odoni_introducing_2019}
F.~Odoni, A.~M. Brasoveanu, P.~Kuntschik, and A.~Weichselbraun, ``Introducing
  orbis: An extendable evaluation pipeline for named entity linking drill-down
  analysis,'' in \emph{82nd Annual Meeting of The Association for Information
  Science (ASIS\&T 2019)}, Melbourne, Australia, October 2019.

\bibitem{DBLP:conf/aclwac/BarbaresiL20}
\BIBentryALTinterwordspacing
A.~Barbaresi and G.~Lejeune, ``Out-of-the-box and into the ditch? multilingual
  evaluation of generic text extraction tools,'' in \emph{Proceedings of the
  12th Web as Corpus Workshop, WAC@LREC 2020, Marseille, France, May 2020},
  A.~Barbaresi, F.~Bildhauer, R.~Sch{\"{a}}fer, and E.~Stemle, Eds.\hskip 1em
  plus 0.5em minus 0.4em\relax European Language Resources Association, 2020,
  pp. 5--13. [Online]. Available:
  \url{https://www.aclweb.org/anthology/2020.wac-1.2/}
\BIBentrySTDinterwordspacing

\bibitem{DBLP:journals/cys/SidoKP19}
\BIBentryALTinterwordspacing
J.~Sido, M.~Konop{\'{\i}}k, and O.~Praz{\'{a}}k, ``English dataset for
  automatic forum extraction,'' \emph{Computaci{\'{o}}n y Sistemas}, vol.~23,
  no.~3, 2019. [Online]. Available:
  \url{https://www.cys.cic.ipn.mx/ojs/index.php/CyS/article/view/3259}
\BIBentrySTDinterwordspacing

\bibitem{levenshtein_binary_1966}
V.~Levenshtein, ``Binary {Codes} {Capable} of {Correcting} {Deletions},
  {Insertions} and {Reversals},'' \emph{Soviet Physics Doklady}, vol.~10, p.
  707, 1966.

\bibitem{weninger_cetr_2010}
T.~Weninger, W.~H. Hsu, and J.~Han, ``{CETR}: content extraction via tag
  ratios,'' in \emph{Proceedings of the 19th {International} {Conference} on
  {World} {Wide} {Web}, {WWW} 2010, {Raleigh}, {North} {Carolina}, {USA},
  {April} 26-30, 2010}, M.~Rappa, P.~Jones, J.~Freire, and S.~Chakrabarti,
  Eds.\hskip 1em plus 0.5em minus 0.4em\relax ACM, 2010, pp. 971--980.

\end{thebibliography}

\end{document}